\PassOptionsToPackage{unicode}{hyperref}
\PassOptionsToPackage{hyphens}{url}
\PassOptionsToPackage{dvipsnames,svgnames,x11names}{xcolor}
\documentclass[
  12pt]{article}

\usepackage{amsmath,amssymb}
\usepackage{iftex}
\ifPDFTeX
  \usepackage[T1]{fontenc}
  \usepackage[utf8]{inputenc}
  \usepackage{textcomp} 
\else 
  \usepackage{unicode-math}
  \defaultfontfeatures{Scale=MatchLowercase}
  \defaultfontfeatures[\rmfamily]{Ligatures=TeX,Scale=1}
\fi
\usepackage{lmodern}
\ifPDFTeX\else  
\fi
\IfFileExists{upquote.sty}{\usepackage{upquote}}{}
\IfFileExists{microtype.sty}{
  \usepackage[]{microtype}
  \UseMicrotypeSet[protrusion]{basicmath} 
}{}
\makeatletter
\@ifundefined{KOMAClassName}{
  \IfFileExists{parskip.sty}{%
    \usepackage{parskip}
  }{
    \setlength{\parindent}{0pt}
    \setlength{\parskip}{6pt plus 2pt minus 1pt}}
}{
  \KOMAoptions{parskip=half}}
\makeatother
\usepackage{xcolor}
\makeatletter
\ifx\paragraph\undefined\else
  \let\oldparagraph\paragraph
  \renewcommand{\paragraph}{
    \@ifstar
      \xxxParagraphStar
      \xxxParagraphNoStar
  }
  \newcommand{\xxxParagraphStar}[1]{\oldparagraph*{#1}\mbox{}}
  \newcommand{\xxxParagraphNoStar}[1]{\oldparagraph{#1}\mbox{}}
\fi
\ifx\subparagraph\undefined\else
  \let\oldsubparagraph\subparagraph
  \renewcommand{\subparagraph}{
    \@ifstar
      \xxxSubParagraphStar
      \xxxSubParagraphNoStar
  }
  \newcommand{\xxxSubParagraphStar}[1]{\oldsubparagraph*{#1}\mbox{}}
  \newcommand{\xxxSubParagraphNoStar}[1]{\oldsubparagraph{#1}\mbox{}}
\fi
\makeatother

\usepackage{longtable,booktabs,array}
\usepackage{calc} 
\usepackage{etoolbox}
\makeatletter
\patchcmd\longtable{\par}{\if@noskipsec\mbox{}\fi\par}{}{}
\makeatother
\IfFileExists{footnotehyper.sty}{\usepackage{footnotehyper}}{\usepackage{footnote}}
\makesavenoteenv{longtable}
\usepackage{graphicx}
\makeatletter
\def\maxwidth{\ifdim\Gin@nat@width>\linewidth\linewidth\else\Gin@nat@width\fi}
\def\maxheight{\ifdim\Gin@nat@height>\textheight\textheight\else\Gin@nat@height\fi}
\makeatother
\setkeys{Gin}{width=\maxwidth,height=\maxheight,keepaspectratio}
\makeatletter
\def\fps@figure{htbp}
\makeatother

\makeatletter
\@ifpackageloaded{caption}{}{\usepackage{caption}}
\AtBeginDocument{%
\ifdefined\contentsname
  \renewcommand*\contentsname{Table of contents}
\else
  \newcommand\contentsname{Table of contents}
\fi
\ifdefined\listfigurename
  \renewcommand*\listfigurename{List of Figures}
\else
  \newcommand\listfigurename{List of Figures}
\fi
\ifdefined\listtablename
  \renewcommand*\listtablename{List of Tables}
\else
  \newcommand\listtablename{List of Tables}
\fi
\ifdefined\figurename
  \renewcommand*\figurename{Figure}
\else
  \newcommand\figurename{Figure}
\fi
\ifdefined\tablename
  \renewcommand*\tablename{Table}
\else
  \newcommand\tablename{Table}
\fi
}
\@ifpackageloaded{float}{}{\usepackage{float}}
\floatstyle{ruled}
\@ifundefined{c@chapter}{\newfloat{codelisting}{h}{lop}}{\newfloat{codelisting}{h}{lop}[chapter]}
\floatname{codelisting}{Listing}

\makeatother
\makeatletter
\@ifpackageloaded{caption}{}{\usepackage{caption}}
\@ifpackageloaded{subcaption}{}{\usepackage{subcaption}}
\makeatother

\ifLuaTeX
  \usepackage{selnolig}  
\fi
\usepackage[]{natbib}
\usepackage{bookmark}

\IfFileExists{xurl.sty}{\usepackage{xurl}}{} 
\hypersetup{
  pdftitle={Title},
  pdfauthor={Author 1; Author 2},
  pdfkeywords={3 to 6 keywords, that do not appear in the title},
  colorlinks=true,
  linkcolor={blue},
  filecolor={Maroon},
  citecolor={Blue},
  urlcolor={Blue},
  pdfcreator={LaTeX via pandoc}}

\usepackage{amsmath,bm}
\usepackage{color,subcaption,multirow,multicol,comment,longtable}
\usepackage{titletoc}
\usepackage{cleveref}
\usepackage[ruled,linesnumbered]{algorithm2e}
\usepackage{array}    
\usepackage{booktabs}
\usepackage{subcaption}
\usepackage{xcolor}

\usepackage{hyperref}

\usepackage{soul}
\usepackage{xurl}

\def\1{\bm{1}}

\newtheorem{theorem}{Theorem}[section]

\newtheorem{proposition}{Proposition}[section]

\newtheorem{assumption}{Assumption}
\newtheorem{definition}{Definition}
\newtheorem{example}{Example}

\newcommand{\anon}{1}

\begin{document}
\addtocontents{toc}{\protect\setcounter{tocdepth}{-1}}

\def\spacingset#1{\renewcommand{\baselinestretch}%
{#1}\small\normalsize} \spacingset{1}


\if1\anon
{
  \title{\bf Deep Ranking with Heterogeneous Effects}
  \author{Yuanhang $\rm{Luo}^{a}$\thanks{Equally contributed authors; ${}^{\dagger}$ Corresponding authors},\hspace{0.2cm}
		Shuxing $\rm{Fang}^{a*}$,\hspace{0.2cm}
		Ruijian $\rm{Han}^{a\dagger}$\hspace{0.2cm} and
		Yiming $\rm{Xu}^{b\dagger}$
		\\
		{\small {\small {$\it^{a}$  Department of Data Science and Artificial Intelligence, The Hong Kong Polytechnic University  } }}\\
		{\small {\small {$\it^{b}$ Department of Mathematics, University of Kentucky} }}
	}
  \maketitle
} \fi

\if0\anon
{
  \bigskip
  \bigskip
  \bigskip
  \begin{center}
    {\LARGE\bf Title}
\end{center}
  \medskip
} \fi

\bigskip
\begin{abstract}
Classical latent-score ranking models often fail to distinguish objects' intrinsic scores from contextual effects, which are typically nonlinear and can dominate the observed outcomes. To address this, we introduce a semiparametric ranking framework in which the log-score of each object is modeled as the sum of a utility parameter and a nonparametric covariate effects. Within this framework, we establish model identifiability under mild regularity and connectivity conditions. For estimation, we approximate the covariate effects using a neural network and estimate the parameters via maximum likelihood. Under random design assumptions, we prove that the resulting estimator exists with high probability and derive non-asymptotic error bounds that achieve minimax optimality for both the parametric and nonparametric components. Numerical experiments on both synthetic data and an ATP tennis dataset are conducted to support our findings.
\end{abstract}

\noindent%
\vfill

\newpage
\spacingset{1.2} 

\section{Introduction}\label{sec:intro}
Ranking data arising from comparisons of a set of $n$ objects is a prevalent task in sports analytics, social science, and machine learning. Many classical latent-score models assign a positive \textit{score} to each object to measure their strength \citep{thurstone1927method, zermelo1929berechnung}. A standard example is the Plackett--Luce (PL) model \citep{luce1959individual, plackett1975analysis}. Consider a comparison involving a subset of objects $e \subseteq [n]$ of size $m=|e|$, where $[n]=\{1,\ldots,n\}$. The outcome is a random permutation $\pi = (\pi(1), \ldots, \pi(m))$, where $\pi(j) \in e$ denotes the object ranked at the $j$th position in $e$. For convenience, we use the log-transform of scores, denoted by $\bm u = (u_1, \ldots, u_n)^\top\in\mathbb R^n$ and referred to as the \textit{utility} vector. The PL model posits that the probability of observing $\pi$ is given by
\begin{equation}
\label{PL}
    \mathbb{P}_{\boldsymbol{u}}\left(\pi \mid e\right)=\prod_{j \in[m]} \frac{\exp (u_{\pi(j)})}{ \exp (u_{\pi(j)})+\cdots + \exp (u_{\pi(m)})}.
\end{equation}
This formulation implicitly assumes a sequential choice sampling process. Specifically, the highest-ranked object is first drawn from the set of 
$m$ candidates proportional to their scores. Subsequent objects are then sequentially selected from the remaining unranked candidates according to the same rule until a full ranking is constructed.

When $m=2$, the PL model reduces to the Bradley--Terry (BT) model \citep{bradley1952rank}. The statistical foundations of both the BT and PL models have been extensively studied in the modern ranking literature \citep{simons1999asymptotics,hunter2004mm, MR2987494,shah2016estimation,Negahban2017rank,chen2019spectral,chen2022partial,gao2023uncertainty,fan2024uncertainty,han2025unifiedanalysis,fan2025spectral}; see \cite{fang2026recent} for a detailed review.

The distribution in \eqref{PL} is determined by the utility vector $\bm u^*$ and remains static conditional on $e$. In many applications, however, repeated comparisons among the same subset of objects are collected over time or under varying contexts.  
These contexts may encompass auxiliary information such as demographic attributes, physical conditions, inventory levels, and task prompts, which can significantly influence the outcomes. The PL formulation in \eqref{PL} does not account for such covariate information and may lead to a loss of predictive power in practice.

To address the limitation of static scores, this paper develops a semiparametric framework that combines the interpretability of the PL model with nonlinear function approximation. Let $X_{e,j} \in \mathbb{R}^d$ denote the contextual feature associated with object $j$ in a comparison $e$ (i.e., $j\in e$ and $X_{e,j}$ is edge-dependent). We parameterize the log-score of object $j$ on hyperedge $e$ as the sum of an intrinsic utility and nonlinear covariate effects:
$$s_{e,j}(\boldsymbol{u}, f)\;=\;u_j + f\!\left(X_{e,j}\right),
\qquad
(\boldsymbol{u}, f)\in \mathbb{R}^n \times \mathcal{F},$$
where $\mathcal{F}$ is some function class (which can be infinite-dimensional) with sufficient expressiveness. Under this formulation, the PL likelihood for a ranking $\pi$ is modified as
\begin{align}\label{dr:formula}
\mathbb{P}_{\boldsymbol{u}, f}\!\left(\pi \,\middle|\, \left\{X_{e,j}\right\}_{j\in e}, e\right)
\;=\;
\prod_{j\in [m]}
\frac{\exp\left( s_{e,\pi(j)}\left(\boldsymbol{u}, f\right) \right)}
{\sum_{t=j}^m \exp\left( s_{e,\pi(t)}\left(\boldsymbol{u}, f\right) \right)}.
\end{align}

Compared to the PL model \eqref{PL}, the formulation in \eqref{dr:formula} separates the utility $u_j$ from the contextual effects $f(X_{e,j})$. This allows us to model outcome probabilities under varying conditions while holding the intrinsic utility fixed 
(e.g., {results in a sprint race may depend on the lane assignment and other ambient conditions}). 
From an axiomatic perspective, this decomposition mitigates violations of stochastic transitivity; see Section~\ref{sec:2} for an extended discussion. 

Throughout this paper, we use  $(\boldsymbol{u}^*, f^*)$ to denote the true underlying parameters that generate the comparison outcomes. In practice, the dimension of $f^*$ can be large, and classical nonparametric methods based on kernel-smoothing or splines become inefficient. To address this, we leverage recent developments in deep learning to approximate $f^*$ using a deep neural network (DNN). This approach enables the estimation of model parameters via the maximum likelihood estimator (MLE) while maintaining the flexibility to capture complex, high-dimensional patterns. Modern machine learning architectures also facilitate efficient computation under this parametrization. In the remainder of this article, we refer to the model in \eqref{dr:formula}, with $\mathcal{F}$ specified as a class of DNNs, as the Deep Heterogeneous Ranking (DHR) model.

\subsection{Related Work} 

Incorporating covariate information into comparison models has been widely studied under parametric assumptions. Existing approaches generally fall into two categories: those that model the log-scores entirely through covariate effects \citep{cattelan2012models,ijcai2018p0304,schafer2018dyad}, and those that augment latent utilities with covariate information \citep{schauberger2019btllasso,li2022bayesian,  fan2024covariate,fan2024uncertainty, dong2025statistical, singh2025least}. Our work aligns with the latter category. Within this scope, the model most relevant to ours is the PL model with dynamic covariates (PlusDC) \citep{dong2025statistical}, which assumes a linear model for the population effects $f$ in \eqref{dr:formula}. In contrast, our model does not assume an oracle parametric form for $f$. This flexibility makes our model better suited to capture the complex nonlinearities in large-scale datasets common in modern machine learning applications \citep{christiano2017deep, ouyang2022training}. 

Our work leverages recent advances in statistical approximation using DNNs.  \citet{schmidt2020nonparametric}, \citet{farrell2021deep}, and \citet{jiao_deepnonparametric} established non-asymptotic error bounds for nonparametric regression using DNNs and showed that these models achieve minimax optimal convergence rates under certain smoothness assumptions. Recently, this framework has been extended to reinforcement learning. For instance, \citet{sun2025rethinking} and \citet{luo2025learningguaranteerewardmodeling} analyzed the BT and general comparison models where the reward function is parameterized by a neural network, a setup common in reinforcement learning with human feedback. However, these works did not consider the intrinsic utilities, which introduce additional heterogeneous effects that need to be addressed using separate tools from ranking data analysis \citep{han2020aoap, han2023general, han2025unifiedanalysis, dong2025statistical}.  

Our work is also related to the semiparametric models \citep{robinson1988root, bickel1993efficient}, where the goal is to 
jointly estimate a finite-dimensional parameter of interest alongside 
an infinite-dimensional nuisance function.   Recent studies, such as \citet{farrell2021deep} and \citet{Hu2025aiasflesh}, have analyzed such models when the nuisance function is approximated by a DNN, establishing consistent estimation of both components. Our setting shares this semiparametric nature, but the growing dimensionality of the utility parameters requires tools that go beyond the classical semiparametric framework.

\subsection{Contributions}

In this article, we introduce a general semiparametric ranking framework based on the PL model. Our main contributions are summarized as follows.
\begin{itemize}
\item We introduce a general semiparametric statistical ranking model in \eqref{dr:formula} to model the log-score of each object as an additive combination of intrinsic utilities and nonlinear covariate effects. The proposed model substantially increases the expressiveness of the model in \cite{dong2025statistical} and allows for accommodating both stochastic intransitivity and nonlinear contextual effects. We establish model identifiability under mild graph connectivity and regularity assumptions. 

\item We parametrize the nonparametric component in \eqref{dr:formula} using a DNN and apply the MLE for parameter estimation. Under additional assumptions, we show the existence of the MLE with high probability and derive non-asymptotic error bounds for both the estimated intrinsic utility parameters and the DNN component. An informal version of our result, specialized to the pairwise comparison setting, is given as follows: 
\paragraph*{Main Result (informal)}
Suppose that $f^*$ is $\beta$-Hölder continuous, and $\bm u^*$ is uniformly bounded. Assume that the comparison graph is sampled from an Erdős--R\'enyi model $\mathcal G(n, p)$ with $p\gtrsim n^{-(1-\alpha)}$ for some $\alpha<1$. Under mild regularity conditions on the distributions of covariates, there exists an explicit DNN architecture under which the MLE $(\widehat{\boldsymbol{u}},\bar{f}_{\widehat{\phi}})$ almost surely satisfies the following bounds 
up to logarithmic factors:
\begin{equation*}
    \|\widehat{\boldsymbol{u}}-\boldsymbol{u}^*\|_{\Lambda} \lesssim N^{-\frac{\beta}{2\beta +d}} + \sqrt{\frac{n}{N}}, \quad \|\bar{f}_{\widehat{\phi}}-f^*\|_{\mathcal{L}^2(\mathcal{X})} \lesssim N^{-\frac{\beta}{2\beta +d}}  + \sqrt{\frac{n}{N}},
    \end{equation*}
    where $N$ is the total number of observed comparisons, $\|\cdot\|_{\Lambda}$ denotes the semi-norm induced by the graph Laplacian of $\mathcal G$, and $\|\cdot\|_{\mathcal{L}^2(\mathcal{X})}$ denotes the standard $\mathcal{L}^2$ norm weighted over the covariate space $\mathcal{X}$. The implicit constant depends on $\alpha$.  
\end{itemize}

The graph-Laplacian seminorm $\|\cdot\|_\Lambda$ codifies the comparison graph topology and was used by \citet{shah2016estimation} to establish minimax error bounds for the MLE in the BT model. In the hypergraph setting, $\|\cdot\|_\Lambda$ corresponds to the graph Laplacian of an associated graph obtained through edge breaking (see \Cref{sec:theory}). In our framework, the error bounds for both the estimated utility and the DNN components match their respective lower bounds in \citet[Theorem 1]{shah2016estimation} and \citet[Theorem 1]{stone1982optimal}.

The main difficulty in establishing this result arises from the growing dimension of the utility parameters and the adaptive increase in the complexity of DNNs for nonparametric estimation. Estimation error from either part is coupled with the other through comparison graph structures. Establishing estimation error bounds for these components simultaneously while maintaining optimality requires building connections between statistical ranking and nonparametric learning theories. The core innovation of our proof is that we generalize the chaining technique introduced by \citet{han2023general} to locate the MLE $\widehat{\bm u}$ within an $\ell_\infty$-neighborhood of $\bm u^*$ via the DNN estimation error. This step provides the foundation for subsequent empirical process analysis to control the errors from both the parametric and nonparametric components.

To the best of our knowledge, this is the first non-asymptotic analysis of a semiparametric framework for covariate-assisted statistical ranking. To support our theoretical findings, we conduct extensive experiments demonstrating that the proposed approach yields empirical gains over standard BT/PL baselines and linear-covariate specifications on both synthetic and real-world datasets. 

\subsection{Notation}

We denote by $\mathbb{N}$ and $\mathbb{R}$ the sets of natural and real numbers, respectively. For $n\in\mathbb{N}$, let $[n]=\{1,\ldots,n\}$. Let $1\leq p\leq\infty$. For a vector $\boldsymbol{v}=(v_1,\ldots,v_n)^\top\in\mathbb{R}^n$, its $\ell_p$-norm is
denoted by $\|\boldsymbol{v}\|_p$. We denote $\mathbf{b}_i \in \mathbb{R}^n$ as the $i$-th standard basis vector.
Given a probability measure $\rho$ on the domain $\mathcal{X}$, its induced $\mathcal{L}^p(\mathcal{X})$-norm is denoted by $\|f\|_{\mathcal{L}^p(\mathcal{X})}$. For sequences $a_n$ and $b_n$, the relation $a_n \lesssim b_n$, or equivalently $a_n = \mathcal{O}(b_n)$, means that there exists a constant $c>0$ (independent of $n$) such that $a_n \le c\,b_n$ for all $n$; we write $a_n \asymp b_n$ when both $a_n \lesssim b_n$ and $b_n \lesssim a_n$ hold. Additionally, we use the notation $a_n = \widetilde{\mathcal{O}}(b_n)$ to indicate that $a_n \lesssim b_n (\log n)^k$ for some absolute constant $k \ge 0$. For a sequence of random variables $X_n$, we write $X_n = \mathcal{O}_p(a_n)$ to indicate that $X_n/a_n$ is bounded in probability.

\section{Problem Set-up}\label{sec:2}
We first introduce a general semiparametric ranking framework, which generalizes existing parametric approaches by incorporating unspecified, nonparametric covariate effects.

Recall that $e$ denotes a set representing a comparison among a subset of the $n$ objects, with $m= |e|$. A permutation $\pi$ of $e$ is a bijection from $[m]$ to $e$, which we represent as the ordered tuple $(\pi(1), \ldots, \pi(m))$, where $\pi(i)$ is the object with rank $i$ under $\pi$. We denote by $\mathsf{Perm}(e)$ the set of all such permutations. It is sometimes convenient to work with ranks directly: for each object $j \in e$, let $r(j) \in[m]$ denote its rank under $\pi$, so that $r(\pi(i))=i$ for all $i \in[m]$.

Each object is associated with an edge-dependent covariate $X_{e, j} \in \mathcal{X}$, where $\mathcal{X}\subset\mathbb R^d$ is assumed compact. The edge-dependent assumption not only makes the model practical, but also plays a role in ensuring the model's identifiability and deriving the non-asymptotic error bound (see also Section~\ref{section:identifiability} and \ref{sec:theory}).  The random variable $X_{e, j}$ can be understood as observable attributes of object $j$ in comparison $e$. Our general semiparametric framework specifies the log-score of object $j$ on edge $e$ as
\begin{align}
s_{e,j}(\boldsymbol{u}, f)=u_j+f\left(X_{e, j}\right), \quad (\boldsymbol{u}, f) \in \mathbb{R}^n \times \mathcal{F},\label{semiparametric}
\end{align}
\noindent where $\boldsymbol{u}=\left(u_1, \ldots, u_n\right)^{\top}$ represents object-specific utility and $f: \mathcal{X} \rightarrow \mathbb{R}$ is a nonparametric term that captures the effects of dynamic covariates. This formulation extends the PlusDC model of \citet{dong2025statistical} by replacing linear covariate effects with a flexible nonparametric component, thereby enabling the capture of nonlinear patterns and higher-order interactions without requiring prior specification of the functional form.

In this setting, the probability of observing a particular permutation $\pi=(\pi(1), \ldots, \pi(m))$ of the objects on hyperedge $e$ follows the PL type of formulation:
\begin{equation}
\label{deep_ranking}
    \mathbb{P}_{\boldsymbol{u}, f}\left(\pi \mid \left\{X_{e, j}\right\}_{j \in e},e\right)=\prod_{j \in[m]} \frac{\exp \left(s_{e,\pi(j)}\left(\boldsymbol{u}, f\right)\right)}{\sum_{t=j}^m \exp \left(s_{e,\pi(t)}\left(\boldsymbol{u}, f\right)\right)}.
\end{equation}
\noindent {As in the standard PL model, our semi-parametric model \eqref{deep_ranking} 
admits the similar sequential choice interpretation, with the score $s_{e,j}(\boldsymbol{u}, f)$ 
now incorporating both the intrinsic utility and the covariate effects.}

A critical advantage of introducing the context-dependent component $f$ is that it enables the model to capture more complex comparison structures, such as cyclic dominance, which standard PL/BT models fail to represent in expectation. 
\begin{example}[Accommodating intransitivity]
Consider a tennis tournament among three players $\{a, b, c\}\subset \mathcal{V}$ with utilities $u^*_a = 1$, $u^*_b = 0$, and $u^*_c = -1$. The performance is determined by players' utilities and their age, $X_{e,j}$. The true age effect $f^*$ is non-monotone, with $f^*(20)=0$, peaking at $f^*(25)=3$, and declining by age 30 with $f^*(30)=0$. We observe outcomes across three specific designs:

\begin{enumerate}
    \item $a$ vs. $b$, at age $X_{1,a}=X_{1,b}=25$. Then  $a\succ b$ is the most probable ranking.
\item $b$ vs. $c$, with $X_{2,b}=25,X_{2,c}=20$. Then $b\succ c$ is the most probable ranking.
 \item $c$ vs. $a$, with $X_{3,c}=25,X_{3,a}=30$. Then  $c\succ a$ is the most probable ranking.
\end{enumerate}
\end{example}

While the BT/PL model may generate cyclic observations (e.g., $a \succ b$, $b \succ c$, $c \succ a$) due to the probabilistic nature, it enforces transitivity in expectation. Since the PlusDC model assumes linear contextual effects—and given that the age difference between players remains constant even as they age—population transitivity is preserved within that framework. In contrast, our semiparametric approach addresses this restriction by allowing the contextual component $f^*(X_{e,j})$ to take a general form. The transitivity property in PL/BT models is fundamentally due to the curl-free condition on the pairwise comparison probability matrix \citep{lim2020hodge}. Alternative methods that mitigate violations of transitivity by relaxing this constraint while using low-rank approximations to preserve essential skew-symmetry can be found in \citet{lee2025pairwisetransitivity}.

The formulation in \eqref{deep_ranking} builds upon the semiparametric \textit{polychotomous choice model} introduced by \citet{matzkin1991semiparametric}. 
Consider a fixed, universal set of alternatives  \(\mathcal{V}=\{1,\ldots,n\}\), with \(X_j\) representing the observable attributes of item \(j \in \mathcal{V}\). The polychotomous choice model is defined based on the following choice mechanism:
\begin{example}[Polychotomous choice model]
\label{matzkin}
Item $j$ is chosen from $\mathcal{V}$ if 
$$f\left(X_j\right)+\varepsilon_j \geq f\left(X_k\right)+\varepsilon_k \quad \text{for all } k \in \mathcal{V}\backslash\{j\},$$
where $\varepsilon_j$ are unobservable error terms and assumed to follow some extreme value distributions according to $u_j$. 
\end{example}
When the error terms $\varepsilon_j$ are assumed to follow independent Gumbel distributions with location parameter $u_j$, the probability of selecting item $j$ as the top-one choice reduces to the softmax form in \eqref{deep_ranking}. Our model also allows comparisons over subsets of items, which is relevant for applications such as sports tournaments. 

\subsection{Comparison Graphs}\label{Random_Graph_Model}

\begin{figure}[!ht]
    \centering

    \begin{subfigure}[b]{0.4\textwidth}
        \centering
        \includegraphics[width=\textwidth]{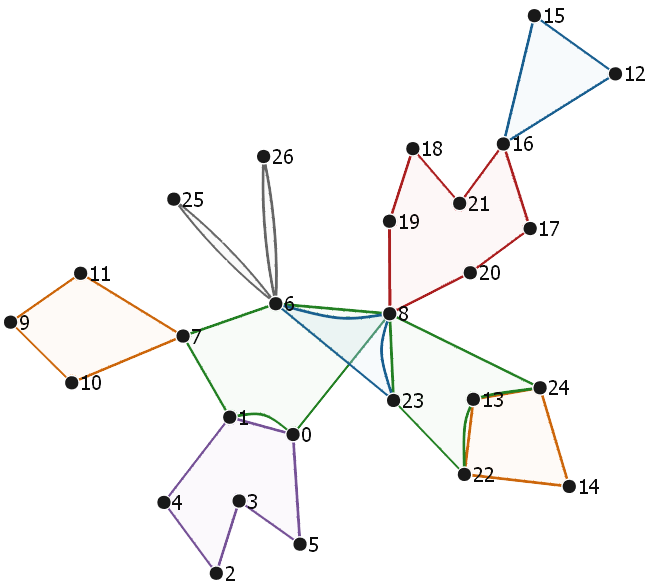}
        \caption{}
        \label{fig:graph1}
    \end{subfigure}
    \begin{subfigure}[b]{0.4\textwidth}
        \centering
        \includegraphics[width=\textwidth]{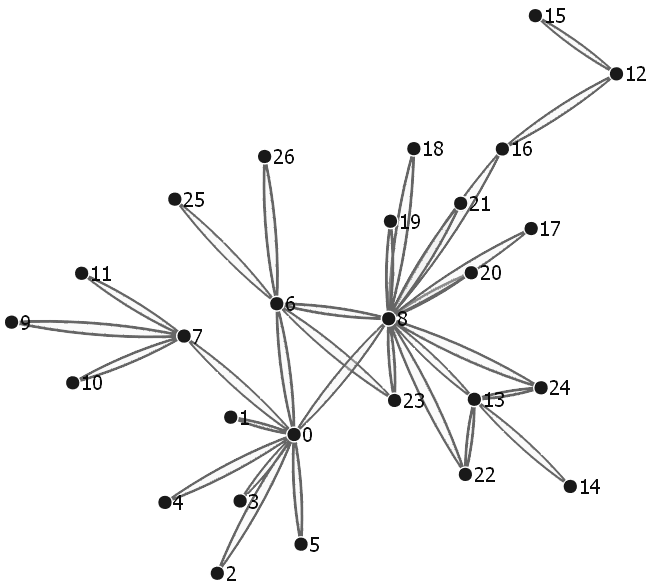}
        \caption{}
        \label{fig:graph2}
    \end{subfigure}
    \caption{Illustration of comparison graph structures\protect\footnotemark. (a) A general hypergraph representation where comparisons involve sets of items (hyperedges) with varying sizes $m \in \{2, \dots, 7\}$. (b) A standard pairwise comparison graph where interactions are restricted to pairs of vertices.}
    \label{fig:graph_vis}
\end{figure}

We represent a comparison data as a hypergraph $\mathcal{H}(\mathcal{V},\mathcal{E})$. The vertex set $\mathcal{V}=[n]$ represents the $n$ items under consideration, and the edge set $\mathcal{E}=\{e_i\}_{i\in [N]}$ consists of $N$ hyperedges, where $e_i$ refers to $i$th hyperedge. An illustration of this data structure is provided in Figure \ref{fig:graph_vis}. In the rest of the article, we assume the size of comparison edges to be uniformly bounded.

 \footnotetext{The hypergraph in Figure \ref{fig:graph_vis} is visualized using HGPolyVis 
\citep{oliver2024structure}.
}
\begin{assumption}\label{bounded_M}
    There exists a universal constant $M$ independent of $n$ such that $|e|\leq M$ for all $e \in \mathcal{E}$.
\end{assumption}

In the context of parameter estimation, we assume the comparison graph $\mathcal{H}(\mathcal{V},\mathcal{E})$ is given and satisfies specific structural properties. To ensure these conditions are met, we follow a common practice in the literature by assuming the graph is drawn from a probabilistic model. This assumption simplifies the discussion by ensuring well-behaved graph topologies, including concentrated degrees and sufficient connectivity, with high probability. Consequently, it serves as a tractable base model for general scenarios.

In this article, we assume that $\mathcal{H}(\mathcal{V},\mathcal{E})$ is generated from the \textit{nonuniform random hypergraph model} (NURHM) \citep{han2025unifiedanalysis}. Specifically, the edge set is constructed as the union of independent layers based on edge size $\mathcal{E}=\bigcup _{m=2}^M \mathcal{E}^{(m)}$. For each size $m$ and any potential hyperedge $e \in\binom{\mathcal{V}}{m}$, we include it based on the indicator $\mathbb{I}_{\{e \in \mathcal{E}^{(m)}\}}$ function which is an independent Bernoulli random variable with probability $p_{e, n}^{(m)}$. To characterize the graph's sparsity, we define the lower and upper bounds of the edge probabilities for each layer $m$ as:
$$
p_n^{(m)} = \min _{e \in\binom{\mathcal{V}}{m}} p_{e, n}^{(m)}  
\quad \text{and} \quad 
q_n^{(m)} = \max _{e \in\binom{\mathcal{V}}{m}} p_{e, n}^{(m)}.
$$
The order of the expected vertex degrees in the graph is bounded between the following quantities:
\begin{equation*}
\xi_{n,-} := \sum_{m=2}^M n^{m-1} p_n^{(m)} 
\quad \text{and} \quad 
\xi_{n,+} := \sum_{m=2}^M n^{m-1} q_n^{(m)}.
\end{equation*}

\noindent Our theoretical analysis assumes $\xi_{n,-} \asymp \xi_{n,+}$, which is a generalized homogeneous condition commonly adopted in the literature \citep{chen2019spectral,han2020aoap,lee2025pairwisetransitivity}. In the special case where $p_n^{(m)}=q_n^{(m)}=p_n^{(M)} \mathbb{I}_{\{m=M\}}$, the NURHM reduces to the classical $M$-way Erdős--R\'enyi model \citep{erdos1960evolution}.

\section{Model Identifiability}
\label{section:identifiability}

In this section, we address the identifiability of the semiparametric model \eqref{deep_ranking}, which contains the PL model as a special instance when $f\equiv 0$. It is well known that the PL model is shift-invariant; that is, the utility vector $\bm{u}$ and $\bm{u} + c\mathbf{1}$ induce the same probability distributions over the edge-wise comparison outcomes for any $c \in \mathbb{R}$. This is typically resolved, for instance, by considering the centered space $\mathbb{U}^n:= \{ \boldsymbol{u} \in \mathbb{R}^n: \mathbf{1}^{\top} \boldsymbol{u} = 0\}$.

However, additional complexities arise when jointly considering the model space  $\mathbb{U}^n \times \mathcal{F}$, which encompasses both the parametric utilities and the nonparametric covariate effects. For example, due to the translation invariance of the softmax function involved in \eqref{deep_ranking}, shifting the function $f$ by a constant $c$ yields an observationally equivalent model, that is,
$$\mathbb{P}_{\boldsymbol{u}, f}\left(\pi \mid \left\{X_{e, j}\right\}_{j \in e},e\right) = \mathbb{P}_{\boldsymbol{u}, f+c}\left(\pi \mid \left\{X_{e, j}\right\}_{j \in e},e\right)$$
\noindent for all edges in $\mathcal{E}$. Thus, additional constraints are needed to resolve this identifiability issue.

To formalize this, we first define model identifiability in this setup. Let \(\mathcal{Z}(e) = \mathsf{Perm}(e) \times \mathcal{X}^{|e|}\) denote the space of observable variables on hyperedge $e$, which includes both the permutation $\pi$ and the associated covariates. Then, the joint density of an observation $z = (\pi, \{X_{e, j}\}_{j\in e}) \in \mathcal{Z}(e)$ is given by
\begin{align*}
g_e(z; \boldsymbol{u}, f) := \rho(\{X_{e, j}\}_{j\in e} \mid e) \cdot \mathbb{P}_{\boldsymbol{u}, f}(\pi \mid \{X_{e, j}\}_{j\in e},e),
\end{align*}
where $\rho(\{X_{e, j}\}_{j\in e} \mid e)$ denotes the probability density function of the covariates on edge $e$. 
\begin{definition}[Model Identifiability]\label{def:identifiability}
   The model \eqref{deep_ranking} is \textit{identifiable} if for any distinct $(\boldsymbol{u}, f) \neq (\boldsymbol{u}', f')$ in $\mathbb{U}^n \times \mathcal{F}$, there exists some $e \in \mathcal{E}$ and a set $Z \subseteq \mathcal{Z}(e)$ with positive measure such that
$$
\int_Z g_e(z; \boldsymbol{u}, f) \, dz \neq \int_Z g_e(z; \boldsymbol{u}', f') \, dz.
$$
\end{definition}

Definition~\ref{def:identifiability} adopts a population-level notion of identifiability, that is, two parameter pairs $(\boldsymbol{u}, f)$ and $(\boldsymbol{u}', f')$ are distinguished whenever they induce different joint densities $g_e$ over the full observable space $\mathcal{Z}(e)$. This is often used in the study of semiparametric models \citep{matzkin1991semiparametric}. Note that this differs from the sample-level identifiability in parametric models \citep{fan2024uncertainty,dong2025statistical}, where any two distinct parameter vectors induce different probabilities over the observed data.

\noindent The following conditions are needed to ensure the model identifiability in our setting. 
\begin{assumption}(Conditions of Identification)
\label{ident_assumption}

\begin{enumerate}
\item[(2.1).] The covariates $X_{e, j}$ are independent and identically distributed random variables with an absolutely continuous measure, denoted by $\rho_{\mathcal{X}}$, that is bounded away from zero and infinity.
\item[(2.2).] For all $f \in \mathcal{F}$, $f$ is continuous on $\mathcal{X}$.
\item[(2.3).] For all $f \in \mathcal{F}$,\; $\mathbb{E}_{\mathcal{X}}[f] = 0$.
\end{enumerate}
\end{assumption}

\noindent Assumption (2.1) ensures sufficient variability and coverage over the covariate space $\mathcal{X}$. The identically distributed assumption is not strictly required, but is adopted to simplify the notation. In particular, $\rho(\{X_{e, j}\}_{j\in e} \mid e) 
= \prod_{j \in e}\rho_{\mathcal{X}}(X_{e,j})$. Assumption (2.2) is a standard continuity assumption on the nonparametric component. Assumption (2.3) centres the covariate effects, modulo common shifts in $f$. 
Under these conditions, we establish the following identifiability result.
\begin{theorem}
\label{lemma_ident}
Let $\mathcal{H}(\mathcal{V}, \mathcal{E})$ be a connected comparison graph. Under Assumption \ref{ident_assumption}, the model \eqref{deep_ranking} is identifiable in the sense of Definition~\ref{def:identifiability}. 
\end{theorem}

\section{Estimation with Deep Neural Networks}

Given a comparison graph $\mathcal{H}(\mathcal{V}, \mathcal{E})$, a collection of covariates $\{X_{e_i, j}\}_{j \in e_i, e_i \in \mathcal{E}}$, and the observed comparison outcomes $\{\pi_i\}_{i \in [N]}$, and denoting $|e_i| = m_i$, the log-likelihood of the tuple $(\boldsymbol{u}, f)$ conditional on $\mathcal{H}(\mathcal{V}, \mathcal{E})$ and $\{X_{e_i, j}\}_{j \in e_i, e_i \in \mathcal{E}}$ is 
\begin{equation}
l(\boldsymbol{u}, f) = \sum_{i \in [N]} \sum_{j \in [m_i]} \left[ s_{ij}(\boldsymbol{u}, f) - \log \left( \sum_{t=j}^{m_i} \exp \left\{ s_{it}(\boldsymbol{u}, f) \right\} \right) \right], \label{eq:mle}
\end{equation}
\noindent where $s_{ij}(\boldsymbol{u}, f) = u_{\pi_i(j)} + f(X_{e_i, \pi_i(j)})$.

In this article, we model the unknown function $f^*$ using a DNN $f(\cdot; \phi)$. Specifically, we employ a standard feedforward architecture with $K$ hidden layers. Let $w_0 = d$ be the input dimension and $w_{K+1} = 1$ be the output dimension. The architecture is defined by a sequence of widths $(w_0, w_1, \ldots, w_{K+1})$. To define the network formally, let $\textsc{ReLU}: \mathbb{R} \to \mathbb{R}$ be a fixed rectified linear unit activation, applied element-wise to vector inputs. We define the $k$th affine transformation $\mathcal{A}_k: \mathbb{R}^{w_{k-1}} \to \mathbb{R}^{w_k}$ as:
\begin{equation}
    \mathcal{A}_k(z) = A^{(k)} z + a^{(k)}, \quad \text{for } k \in [K+1],
\end{equation}
where $A^{(k)} \in \mathbb{R}^{w_k \times w_{k-1}}$ is the weight matrix and $a^{(k)} \in \mathbb{R}^{w_k}$ is the bias vector. The computation of the network propagates through the hidden layers recursively. Let $h^{(0)} = X$. The hidden features $h^{(k)}$ are generated as
$h^{(k)} = \textsc{ReLU} ( \mathcal{A}_k( h^{(k-1)} ) ),$ for $ k \in [K]$. The final output of the network, $f(x; \phi)$, is obtained by applying the final affine transformation to the last hidden layer:
\begin{equation}
\label{DNN_functional}
    f(x; \phi) = \mathcal{A}_{K+1}(h^{(K)}) = A^{(K+1)} \left( \textsc{ReLU} \circ \mathcal{A}_K \circ \cdots \circ \textsc{ReLU} \circ \mathcal{A}_1 (x) \right) + a^{(K+1)}.
\end{equation}
The full set of trainable parameters is denoted by the flattened vector $\phi := \mathsf{vec}( \{ A^{(k)}, a^{(k)} \}_{k=1}^{K+1} ) \in \mathbb{R}^S,$ where $S = \sum_{k=0}^{K} (w_k \times w_{k+1} + w_{k+1})$ represents the total number of parameters. Denoting the maximum width of the neural network as $W:=\max_{k\in[K]} w_k$, the overall complexity of the network (i.e., the number of parameters) satisfies $S = \mathcal{O}(W^2 K)$.

For analysis, we restrict the parameters to a compact set. Let $\mathcal{F}_{\mathsf{DNN}}$ denote the space of such networks with bounded weights:
$$\mathcal{F}_{\mathsf{DNN}} = \left\{ f(\cdot; \phi) \text{ specified as in } \eqref{DNN_functional}: \phi \in [-R_\phi, R_\phi]^S , \|f\|_{\mathcal{L}^\infty(\mathcal{X})}\leq R_f\right\}, $$
{for some $R_\phi \geq 1$ that depends on $N$ and some universal constant $R_f$.} 
Although $\mathcal{F}_{\mathsf{DNN}}$ depends on the network configurations, such as $W$, $K$, and $R_\phi$, we suppress these dependencies in the sequel for notational clarity. We maximize the log-likelihood over the estimating class $\boldsymbol{\Theta} := \left\{ \boldsymbol{u}, f_\phi : \boldsymbol{u} \in \mathbb{U}^n, \ f_\phi \in \mathcal{F}_{\mathsf{DNN}} \right\}$.
\noindent In what follows, we use the shorthand $f_\phi$ to denote neural networks parameterized in this way. The estimation problem is then formulated as
\begin{equation}
\label{mle}
(\widehat{\boldsymbol{u}}, f_{\widehat{\phi}}) \in \underset{(\boldsymbol{u}, f_\phi) \in \boldsymbol{\Theta}}{\operatorname{argmax}} \; l(\boldsymbol{u}, f_\phi).
\end{equation}
To enforce condition~(2.3) in Assumption~\ref{ident_assumption}, we center of $f_{\widehat{\phi}}$ by its empirical mean over the observed covariates to obtain
\begin{equation}\label{dnn_centering}
\bar{f}_{\widehat{\phi}}(x):=f_{\widehat{\phi}}(x)-\frac{1}{N_{\operatorname{obs}}}\sum_{i\in [N]}\sum_{j\in e_i}f_{\widehat{\phi}}(X_{e_i,j}), \quad\quad N_{\operatorname{obs}}:=\sum_{i=1}^N m_i.\end{equation}
We refer to $(\widehat{\boldsymbol{u}}, \bar{f}_{\widehat{\phi}})$ as the MLE. A key challenge in maximum likelihood estimation for ranking models is that the likelihood can become unbounded if the comparison outcomes are not strongly connected (i.e., there exists one block that defeats the other based on the observed outcomes). To ensure the existence of a finite solution, we require the following regularity condition.

\begin{assumption}
\label{both_win_lose}
Given the comparison graph $\mathcal{H} (\mathcal{V}, \mathcal{E})$ and outcomes $\{\pi_e\}_{e \in \mathcal{E}}$, for any vector $\boldsymbol{u} \in \mathbb{U}^n$ and $\boldsymbol{u}\neq \boldsymbol{0}$, there exist objects $k_1, k_2 \in \mathcal{V}$ and a comparison $e \in \mathcal{E}$ such that $k_1 \succ k_2$ in the outcome $\pi_e$, yet $u_{k_1} < u_{k_2}$.
\end{assumption}

\begin{theorem}[Existence of the Solution]
\label{theorem_existence}
The MLE (\ref{mle}) has a solution $(\widehat{\boldsymbol{u}}, f_{\widehat{\phi}})$ if and only if Assumption $\ref{both_win_lose}$ holds.
\end{theorem}

Assumption \ref{both_win_lose} can be viewed as a generalized ``strong connectivity'' condition on the comparison hypergraph, akin to the necessary and sufficient conditions established for the BT model \citep{zermelo1929berechnung, ford1957solution}. In our semiparametric setting, this assumption prevents the divergence of the estimated utility vector $\widehat{\boldsymbol{u}}$ caused by perfect data separation, while the compactness of the space of $\phi$ prevents the divergence in the estimation of $f^*$. These conditions ensure that the likelihood function is bounded over the feasible set, which guarantees the existence of the maximum likelihood estimator without requiring additional constraints on the feature space and regularity information of $\mathcal{F}$.

It should be noted that the MLE's existence is a probabilistic event that depends on the comparison outcomes. To establish that our model can be consistently estimated, we will demonstrate in the sequel that Assumption \ref{both_win_lose} holds with high probability conditional upon suitable comparison graph structures (e.g., NURHM).

We conclude this section by introducing an algorithm to compute the MLE in \eqref{mle}, assuming its existence. Because the neural network parametrization 
$f(\cdot; \phi)$ renders the optimization problem nonconvex, we employ a projected stochastic gradient descent (Projected SGD) approach, as detailed in Algorithm \ref{alg:ranking_sgd}.
This algorithm performs a joint optimization over $\boldsymbol{\Theta}$, updating the utility vector $\boldsymbol{u}$ and the network parameters $\phi$ simultaneously. To enforce the identifiability constraint $\mathbf{1}^\top \boldsymbol{u} = 0$, we apply a projection step to the vector $\boldsymbol{u}$ after every gradient step. While Algorithm \ref{alg:ranking_sgd} describes the general SGD procedure, the specific parameter updating rule is agnostic to the choice of optimizer, and can be substituted with others, such as Adam \citep{kingma2014adam}, in actual implementation.

\begin{algorithm}[htb!]
\caption{Parameter Estimation in DHR using Projected SGD}
\label{alg:ranking_sgd}
\SetAlgoLined
\SetAlgoNoEnd 
\SetKwInOut{Input}{Input}
\SetKwInOut{Initialize}{Initialize}
\SetKwInOut{Output}{Output}
\SetKwFor{For}{for}{do}{end for}
\SetKw{Return}{Return}
\Input{
Hypergraph $\mathcal{H}(\mathcal{V}, \mathcal{E})$ with covariates $\{X_{e_i,j}\}$, learning rates $\eta_u, \eta_\phi$, number of epochs $T$, $N_{\operatorname{obs}}=\sum_{i=1}^N m_i$, initialization $(\boldsymbol{u}^{(0)}, f_{\phi^{(0)}}) \in \boldsymbol{\Theta}$.
}
\For{$t = 1, \ldots, T$}{
    Generate batches $\{\mathcal{B}_1, \dots, \mathcal{B}_Z\} \subseteq [N]$, and set $(\boldsymbol{u}^{(t)}, \phi^{(t)})=(\boldsymbol{u}^{(t-1)}, \phi^{(t-1)})$;\\
    \For{$z = 1, \ldots, Z$}{
        Compute stochastic gradients:
        $$
\boldsymbol{g}_{u,\phi}  = \sum_{i \in \mathcal{B}_z} \sum_{j \in\left[m_i-1\right]} \frac{\sum_{k=j+1}^{m_i} \exp \left(s_{i k}-s_{i j}\right)\nabla_{(i, j, k)}}{1+\sum_{k=j+1}^{m_i} \exp \left(s_{i k}-s_{i j}\right)},$$
        where $s_{ik} = u_{\pi_i(k)}^{(t)} + f_{\phi^{(t)}}(X_{e_i, \pi_i(k)})$ and $$\nabla_{(i, j, k)}= \begin{bmatrix}
\mathbf{b}_{\pi_i(k)}-\mathbf{b}_{\pi_i(j)} \\
\nabla_\phi f_{\phi^{(t)}}\left(X_{e_i, \pi_i(k)}\right)-\nabla_\phi f_{\phi^{(t)}}\left(X_{e_i, \pi_i(j)}\right)
\end{bmatrix};$$
\noindent Update parameters: $(\boldsymbol{u}^{(t)}, \phi^{(t)})\leftarrow (\boldsymbol{u}^{(t)},\phi^{(t)} ) + (\eta_u,\eta_\phi)\boldsymbol{g}_{u,\phi}$; \\

        Project to constraint space: $\boldsymbol{u}^{(t)}\leftarrow \boldsymbol{u}^{(t)} - \frac{1}{n}\boldsymbol{1}^\top \boldsymbol{u}^{(t)} \boldsymbol{1};$
    }
}
Empirical centering: Obtain $\bar{f}_{\phi^{(T)}}$ by applying \eqref{dnn_centering} to $f_{\phi^{(T)}}$;

\Return{$\boldsymbol{u}^{(T)}, \bar{f}_{\phi^{(T)}}$}
\end{algorithm}

\section{Estimation Error Analysis}\label{sec:theory}

In this section, we establish non-asymptotic error bounds for the MLE when the comparison graph is drawn from an NURHM. 
Because the parametric component $\boldsymbol{u}^*\in\mathbb{U}^n$ grows in 
dimension with the number of items $n$, while the nonparametric component $f^*$ is simultaneously approximated by a DNN of increasing complexity, establishing consistency requires a joint analysis that accounts for both the hypergraph topology and the approximation capacity of the estimating class $\mathcal{F}_{\mathsf{DNN}}$. 

The estimation error of the DNN estimator $\bar{f}_{\widehat{\phi}}$ is evaluated in the standard $\mathcal{L}^2(\mathcal{X})$ norm. For the utility vector $\widehat{\boldsymbol{u}}$, we introduce a semi-norm on $\mathbb{U}^n$ that reflects the structure of the comparison hypergraph. To define it, we represent each hyperedge as an ordered tuple representation. Write $e_i = (j_{i1},\ldots,j_{im_i})$ with $j_{i1}<\cdots<j_{im_i}$, independent of the observed ranking. A \emph{pivotal breaking} of $e_i$ pairs the smallest-indexed vertex with every other vertex in $e_i$, inducing a pairwise graph $\mathcal{H}^\dagger = (\mathcal{V},\mathcal{E}^\dagger)$ with
$$\mathcal{E}^{\dagger} = \left\{(j_{i 1}, j_{i t}) : t \in \{2, \dots, m_i\}, \, i \in [N] \right\}.$$
This pivotal breaking generates $N^\dagger = \sum_{i=1}^N(m_i-1)$ pairwise comparisons in total; see Figure~\ref{fig:graph_vis}(\subref{fig:graph2}) for an illustration. We then construct an incidence matrix $\boldsymbol{Q}_n\in\mathbb{R}^{n\times N^\dagger}$ by setting $\boldsymbol{Q}_n[j_{i1},k]=-1$, $\boldsymbol{Q}_n[j_{it},k]=1$ for the $k$th broken edge $(j_{i1},j_{it})$, and zero elsewhere. The semi-norm on $\mathbb{U}^n$ is defined as
\begin{align}\label{uerror}
    \|\boldsymbol{u}-\boldsymbol{u}'\|_{\Lambda_{\boldsymbol{Q}}} := \sqrt{(\boldsymbol{u}-\boldsymbol{u}')^\top\boldsymbol{Q}_n\boldsymbol{Q}_n^\top(\boldsymbol{u}-\boldsymbol{u}')/N}.
\end{align}
The matrix $\boldsymbol{Q}_n\boldsymbol{Q}_n^\top$ encodes the comparison topology and weights estimation errors by the frequency of observed comparisons. It will degenerate to the standard graph Laplacian adopted in \cite{shah2016estimation} when $m_i=2$ for all $i\in[N]$. In particular, if $n_{ij}$ denotes the number of pairwise comparisons between items $i$ and $j$, the squared error takes the explicit form $\sum_{1 \leq i < j \leq n}n_{i j} ( (\boldsymbol{u}_i - \boldsymbol{u}_j) - (\boldsymbol{u}'_i - \boldsymbol{u}'_j) )^2/N$.
Consistency of the utility parameter MLE is measured using the seminorm error defined in \eqref{uerror}.

We now state the regularity conditions required for nonlinear approximation. To control the approximation bias resulting from DNNs, we require $\mathcal{F}$ to belong to a H\"older smoothness class.

\begin{definition}\label{holder_class} (Hölder Function Class). For a given $\beta>0$ and a finite constant $ c_{\mathsf{H}}>0$, and a compact domain $\mathcal{X} \subset \mathbb{R}^d$, the Hölder function class $\mathcal{C}^\beta\left(\mathcal{X}, c_{\mathsf{H}}\right)$ consists of all $f: \mathcal{X} \rightarrow \mathbb{R}$ such that 
\begin{equation}\label{holder_condition}
\max _{\|\boldsymbol{\omega}\|_1 \leq\lfloor\beta\rfloor}\left\|\partial^\omega f\right\|_{\infty} \leq c_{\mathsf{H}}, \quad\quad\text{and}\quad\quad \max _{\|\boldsymbol{\omega}\|_1=\lfloor\beta\rfloor} \sup _{X \neq X^{\prime}} \frac{\left|\partial^\omega f(X)-\partial^\omega f\left(X^{\prime}\right)\right|}{\left\|X-X^{\prime}\right\|_2^{\beta-\lfloor\beta\rfloor}} \leq c_{\mathsf{H}},\end{equation} 
\noindent where $\boldsymbol{\omega}=\left(\omega_1, \ldots, \omega_d\right)^{\top}\in\mathbb N^d$ is a vector of non-negative integers, and $\partial^\omega=\partial^{\omega_1} \ldots \partial^{\omega_d}$ denotes the corresponding partial derivative operator.
\end{definition}
The left-hand inequality in \eqref{holder_condition} requires all partial derivatives of $f$ up 
to order $\lfloor\beta\rfloor$ to exist and be uniformly bounded on 
$\mathcal{X}$. This also includes some non-differentiable functions, \textit{e.g.}, $\beta<1$. The second condition requires every $\lfloor\beta\rfloor$th order partial derivative to be $(\beta-\lfloor\beta\rfloor)$-H\"older continuous with constant $c_{\mathsf{H}}$. Together, functions in 
$\mathcal{C}^\beta(\mathcal{X}, c_{\mathsf{H}})$ are 
smooth enough to be locally approximated by degree-$\lfloor\beta
\rfloor$ polynomials with a controlled remainder.

\begin{assumption}\label{holder_smooth}
    The function space $\mathcal F$ belongs to $\mathcal{C}^\beta(\mathcal{X}, c_{\mathsf{H}})$ for a given $\beta>0$ and a finite constant $0<c_{\mathsf{H}}\leq R_f$.
\end{assumption}

This assumption is standard in the DNN estimation literature \citep{schmidt2020nonparametric, farrell2021deep, jiao_deepnonparametric}. Although the smoothness parameter $\beta$ may not be known \textit{a priori}, the H\"older class encompasses a rich family of regular functions, and the DNN estimator often adapts well to $\beta$. 
The remaining two conditions concern the boundedness of the true model components and the density of the comparison graph.

\begin{assumption}
\label{uniform_bounded}
There exists a universal constant $R < \infty$ such that 
$$\sup _{n \in \mathbb{N}} \max \left\{\left\|\boldsymbol{u}^*\right\|_{\infty}, \|f^*\|_{\mathcal{L}^\infty(\mathcal{X})}\right\} \leq R.$$
\end{assumption}

\begin{assumption}\label{NURHM_sparsity}
For the NURHM defined in Section \ref{Random_Graph_Model}, we assume that $\xi_{n,-}\asymp \xi_{n,+}$, and there exists $0<\alpha<1$, such that $\xi_{n,-}\gtrsim  n^{1-\alpha}$.
\end{assumption}

The sparsity condition $\xi_{n,-} \gtrsim n^{1-\alpha}$  is stronger than the minimum connectivity needed for the asymptotic analysis in the BT/PL models \citep{chen2022partial, han2025unifiedanalysis}, which grows polylogarithmically, i.e., $\xi_{n,-} \gtrsim \operatorname{poly}(\log n)$. This strengthened condition mainly helps us localize the MLE for the utilities within an $\ell_\infty$-neighborhood of $\bm u^*$ with a uniformly bounded radius (independent of $n$) when general covariate effects $f^*$ are present.  Notably, such a challenge does not arise in classical BT/PL analyses due to the absence of covariate information.

\begin{proposition}
\label{lemma_uni_bounded}
Under Assumptions \ref{bounded_M}, \ref{ident_assumption}, \ref{uniform_bounded} and \ref{NURHM_sparsity}, with probability at least $1-n^{-2}$, the MLE exists under NURHM, and there exists an absolute constant $C>0$ such that $\left\|\widehat{\boldsymbol{u}}-\boldsymbol{u}^*\right\|_{\infty} \leq C$.
\end{proposition}
The proof of Proposition~\ref{lemma_uni_bounded} employs a chaining technique to bound $\left\|\widehat{\boldsymbol{u}}-\boldsymbol{u}^*\right\|_{\infty}$ via the diameter of admissible sequences. In the homogeneous setting, we show that the diameter of admissible sequences is bounded by a uniform constant under the strengthened sparsity condition in Assumption \ref{NURHM_sparsity}. Given that the diameter of admissible sequences generalizes the classical concept of graph diameter \citep{han2025unifiedanalysis}, this result may be viewed as an analogue of the diameter asymptotics in the Erdős--R\'enyi setting \cite[Chapter 7]{bollobas2011random} and may be of independent interest. Conceptually, \Cref{lemma_uni_bounded} parallels the localization step in the leave-one-out analysis for the BT model \citep{chen2022partial}.

Proposition~\ref{lemma_uni_bounded} allows us to apply a Taylor expansion to the log-likelihood function to effectively control the estimation errors and to bound the covering numbers arising from the DNN approximation. A synthesis of these analyses arrives at the following non-asymptotic error bounds for the MLE.

\begin{theorem}[Non-asymptotic Estimation Error]\label{thm:consistency}
Under setting of Proposition \ref{lemma_uni_bounded}, and Assumptions \ref{holder_smooth}, with the choice of neural network parameters as 
$W =114(\lfloor\beta\rfloor+1)^2 d^{\lfloor\beta\rfloor+1} $, $K = 21(\lfloor\beta\rfloor+1)^2\lceil N^{d/(2d+4\beta)} \log _2(8 N^{d/(2d+4\beta)})\rceil$, and $R_\phi=N^{2d/(4\beta+2d)}$, for all sufficiently large $n$, the following results hold a.s.,
\begin{equation}\label{non-asymptotic_error}
\begin{aligned}
    &\|\widehat{\boldsymbol{u}}-\boldsymbol{u}^*\|_{\Lambda_{\boldsymbol{Q}}} \lesssim N^{-\frac{\beta}{2\beta +d}} (\log N)^2 + \sqrt{\frac{n}{N}} (\log N)^2,\\
    &\|\bar{f}_{\widehat{\phi}}-f^*\|_{\mathcal{L}^2(\mathcal{X})} \lesssim N^{-\frac{\beta}{2\beta +d}} (\log N)^2 + \sqrt{\frac{n}{N}} (\log N)^2. 
\end{aligned}\end{equation}
\end{theorem}

The bound in~\eqref{non-asymptotic_error} decomposes into two terms reflecting distinct sources of estimation error. The first term, $N^{-\beta/(2\beta+d)}(\log N)^2$, corresponds to the nonparametric rate for estimating $f^*$, governed by the covariate dimension $d$ and the smoothness $\beta$. Up to logarithmic factors, this matches the minimax optimal rate for nonparametric estimation \cite[Theorem 1]{stone1982optimal} and that of using DNN approximation \cite[Theorem 1 \& 3]{schmidt2020nonparametric}. 

\begin{figure}
    \centering
    \includegraphics[width=0.7\linewidth]{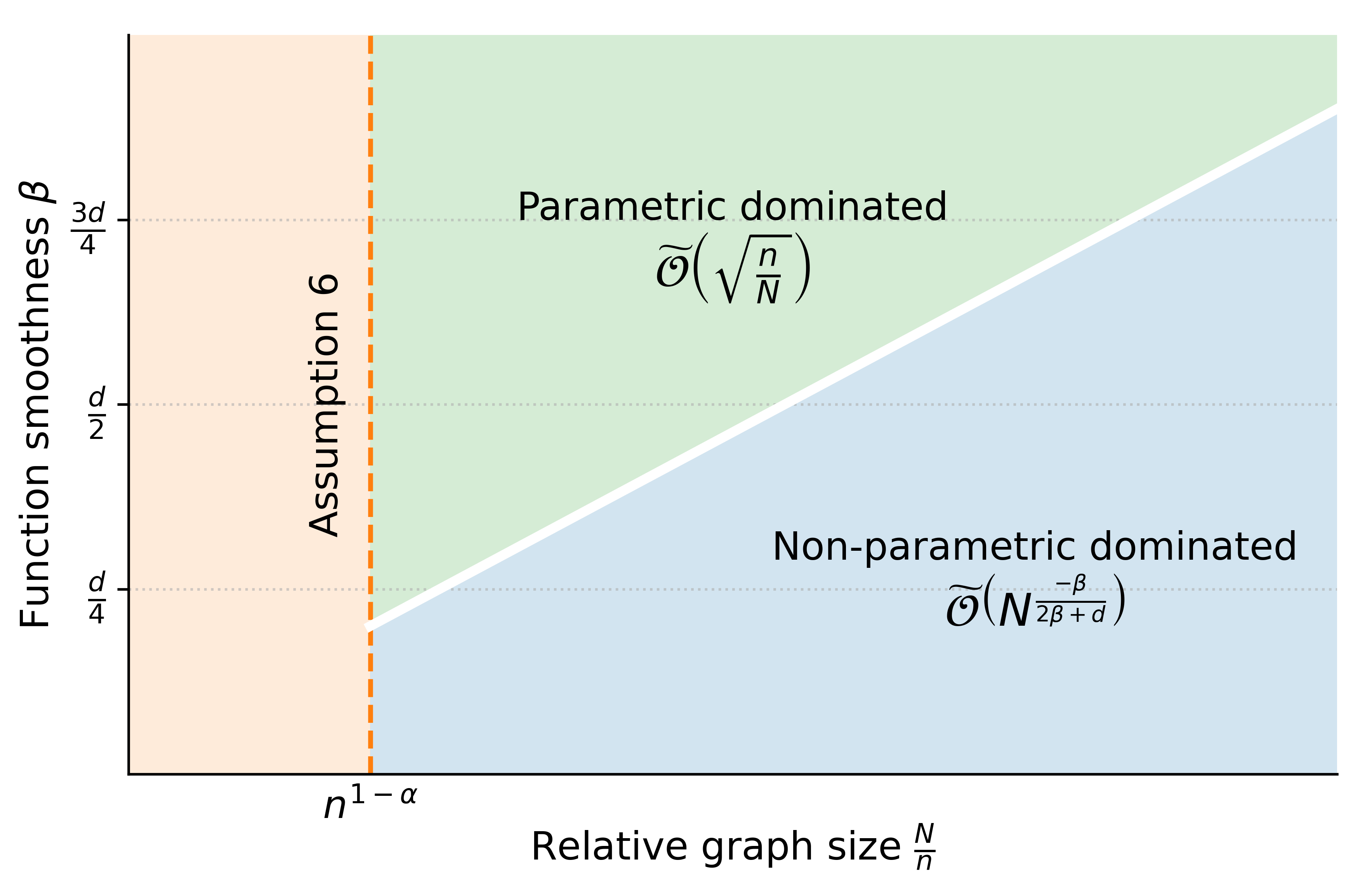}
    \caption{Regimes of error dominance. The parameter space is divided into two regions based on the relationship between function smoothness $\beta$ and the relative sample size $N/n$.
    }
    \label{fig:two_regime}
\end{figure}

The second term, $\sqrt{n/N}(\log N)^2$, corresponds to the parametric rate for estimating the $n$-dimensional utility vector $\boldsymbol{u}^*$ and scales as the square root of the ratio of parameters to sample size. When all comparisons are pairwise ($m_i=2$), this bound matches the minimax optimal rate \cite[Theorem 1]{shah2016estimation}, which gives that for any estimator $\widetilde{\boldsymbol{u}}$ based on $N$ observations from the BT model,
\begin{equation*}\label{eq:lower_parametric}
  \inf_{\widetilde{\boldsymbol{u}}}\;\sup_{\boldsymbol{u}^*\in \{\boldsymbol{u} \in \mathbb{U}^n :\|\boldsymbol{u}\|_\infty \leq R\}}
  \mathbb{E}\!\left[\|\widetilde{\boldsymbol{u}}-\boldsymbol{u}^*
  \|_{\Lambda_{\boldsymbol{Q}}}^2\right]
  \;\gtrsim\; \frac{n}{N},
\end{equation*}
where the expectation is taken with respect to the randomness in the comparison outcomes. In practice, the total estimation error is governed by the slower of these two rates. When the sample size is sufficiently large, i.e., $N \gtrsim n^{1+2 \beta/d}$, the nonparametric error dominates and the total error scales as $\widetilde{\mathcal{O}}(N^{-\beta/(2\beta +d)})$. Conversely, when $N \lesssim n^{1+2 \beta/d}$, the parametric rate dominates, resulting in an overall convergence rate of $\widetilde{\mathcal{O}}(\sqrt{n/N})$. This is illustrated in Figure~\ref{fig:two_regime} and delineates the regime in which the estimation bottleneck shifts from function approximation to parameter estimation.

The architectural parameters $W$ and $K$ specified in Theorem~\ref{thm:consistency} offer qualitative design guidance. When the dimension dominates the smoothness ($d \gg \beta$), the depth simplifies to $K = \widetilde{\mathcal{O}}(\sqrt{N})$ while the width remains polynomial in $d$, suggesting that width primarily addresses input dimensionality whereas depth scales with sample complexity to resolve finer structure in $f^*$. We also emphasize that the error bounds in~\eqref{non-asymptotic_error} characterize the statistical accuracy of the global maximizer of~\eqref{mle}. Given that the objective function involves neural network parametrization and is inherently non-convex, gradient-based optimization methods are typically restricted to identifying local optima in practice. The bounds thus serve as a theoretical baseline for the best achievable statistical performance rather than a guarantee for any specific optimization procedure.

\section{Numerical Study}

In this section, we present numerical experiments on both simulated and real-world datasets. Each dataset is partitioned into training, validation, and test sets. The training and validation sets are used for hyperparameter tuning, while the test set is reserved for evaluating out-of-sample performance. Using simulated data, we empirically verify the uniform consistency of the MLE established in Theorem~\ref{thm:consistency}. We then analyze an ATP tennis dataset to demonstrate the effectiveness of the proposed DHR in capturing complex and nonlinear covariate effects in sports analytics. All experiments were conducted on an Intel Xeon Gold 6330 CPU with 112 cores.

\subsection{Simulations}

To verify the consistency of the MLE, we generate comparison graphs using NURHM. The edge sizes $m_i$ are randomly chosen between $2$ and $8$. The total size of the edge is set to $N=\,5\times n^{1+(1-\alpha)}$. The true intrinsic utility $\boldsymbol{u}^*$ is generated from $\operatorname{Uniform}(-3,3)$ and followed by centering. We draw i.i.d. covariates from $\operatorname{Uniform}(-1, 1)^d$, with $d=2$. For simulation studies, observed matches are split into training and validation sets at an 80/20 ratio. We compute $\|\widehat{\boldsymbol{u}} - \boldsymbol{u}^*\|_{\Lambda_{\boldsymbol{Q}}}$ and $\|\bar{f}_{\widehat{\phi}} - f^*\|_{\mathcal{L}^2(\mathcal{X})}$, with the latter estimated via Monte Carlo integration over the data distribution. All simulation results are averaged over 300 independent replications.

\subsubsection{Varying Relative Graph Size} \label{subsubsec:sim1}
We consider the true function of the following form
\begin{equation}\label{sim1}
    f^*(x_1,x_2)=2\sin(2\pi x_1)\sin(2\pi x_2)+0.5(\exp(x_1^2+x_2^2)-\bar{x}),
\end{equation}
 where $\bar{x}$ is a fixed centering constant. We first examine the convergence property under different graph densities by specifying $(1-\alpha) \in \{0.05, 0.1, 0.15, 0.2, 0.25\}$. The neural network depth is set in proportion to $(\lfloor\beta\rfloor+1)^2\lceil N^{d/(2d+4\beta)} \log _2(8 N^{d/(2d+4\beta)})\rceil$. Note that the function \eqref{sim1} belongs to the $\beta$-Hölder class of any $\beta>0$. For convenience, we set $\beta=1.8$.

\begin{figure}[!ht]
    \centering
    \includegraphics[width=\linewidth]{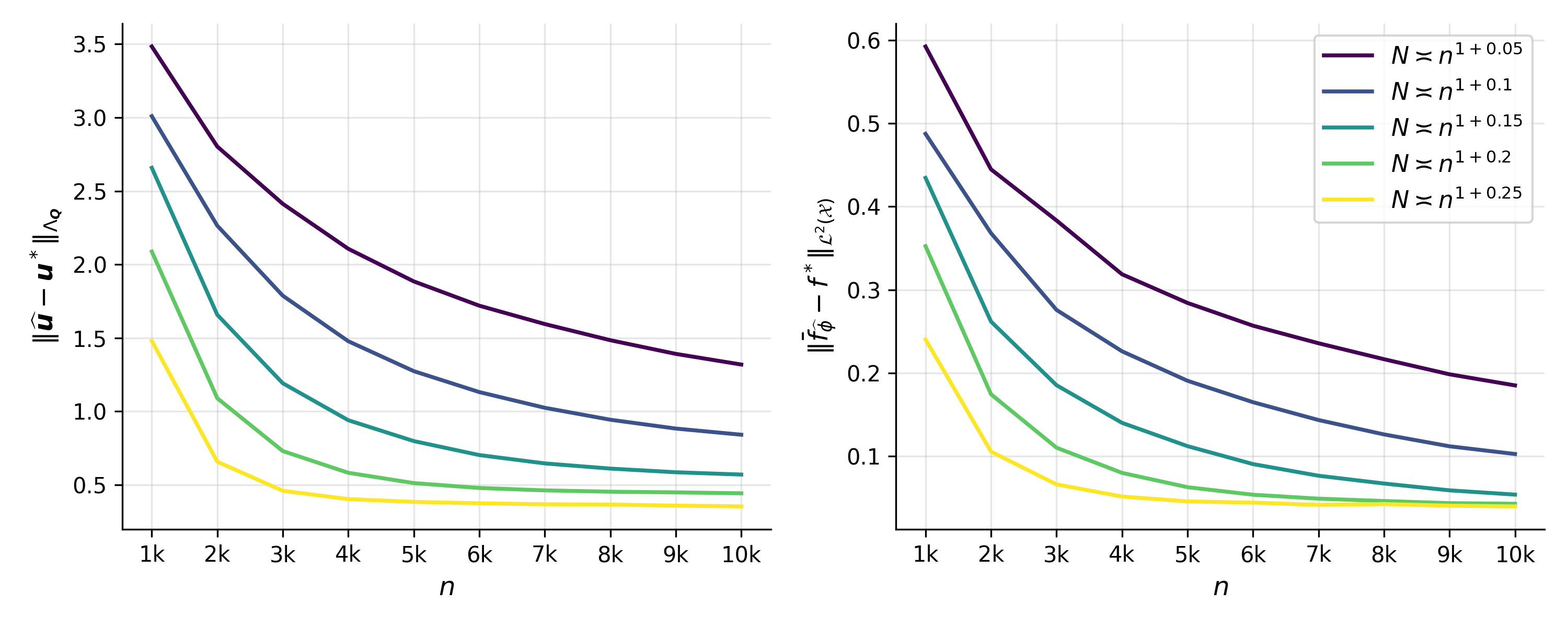}
    \caption{Trace plot of $\| \widehat{\boldsymbol{u}} -\boldsymbol{u}^{*} \|_{\Lambda_{\boldsymbol{Q}}}$ and 
    $\|\bar{f}_{\widehat{\phi}}-f^{*} \|_{\mathcal{L}^{2}(\mathcal{X})}$ as $n$ growing under different graph density regimes. Results are averaged over $300$ replications.}
    \label{fig:TDC}
\end{figure}

\begin{figure}[!ht]
    \centering
\includegraphics[width=\textwidth]{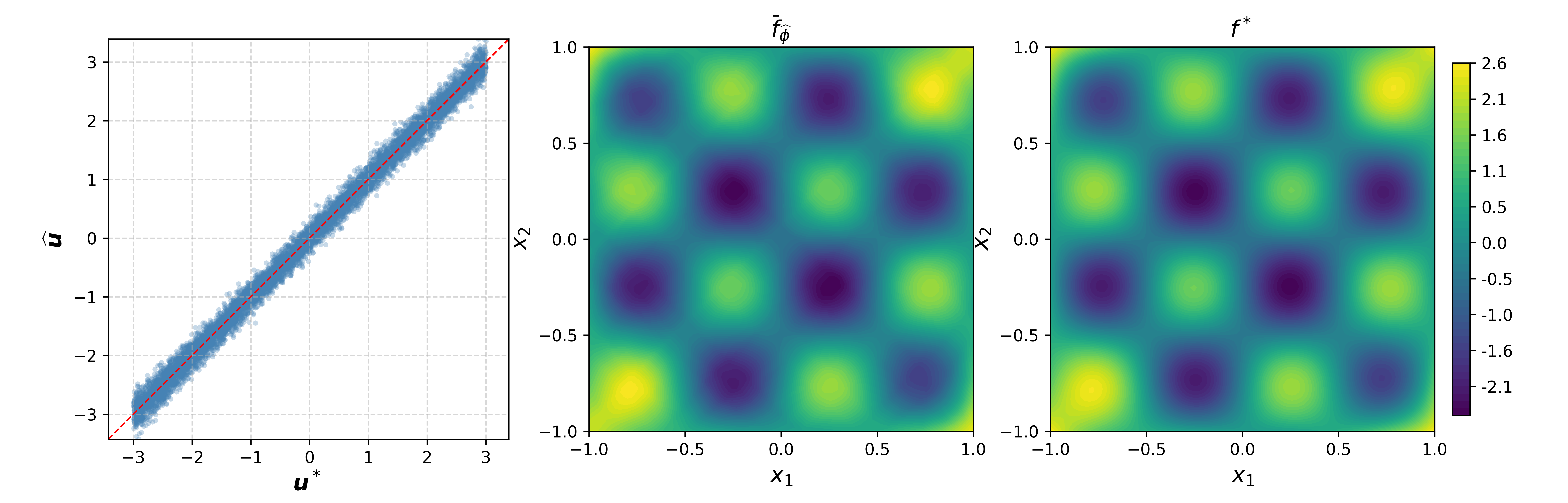}
    \caption{(Left) Scatter plot of the estimated utility $\widehat{\boldsymbol{u}}$ against the ground truth $\boldsymbol{u}^*$. (Middle \& Right) Surface map of $\bar{f}_{\widehat{\phi}}$ and $f^*$ over the domain.}
    \label{fig:two-dim-covariates}
\end{figure}

The results in Figure~\ref{fig:TDC} show the estimation errors of both $\boldsymbol{u}^*$ and $f^*$. It can be observed that for fixed $n$, decreasing $\alpha$ (or making the graph denser) helps accelerate convergence, which is consistent with the monotonic decay of both error terms in \eqref{non-asymptotic_error} with respect to $n/N$. See Figure \ref{fig:two-dim-covariates} for an instance of estimated utility $\boldsymbol{u}^*$ and target function $f^*$ in the setting $n=10,000$ and $(1-\alpha)=0.25$.

\subsubsection{Varying Function Smoothness} \label{simumlationB}
\begin{figure}[!ht]
    \centering
    \includegraphics[width=\linewidth]{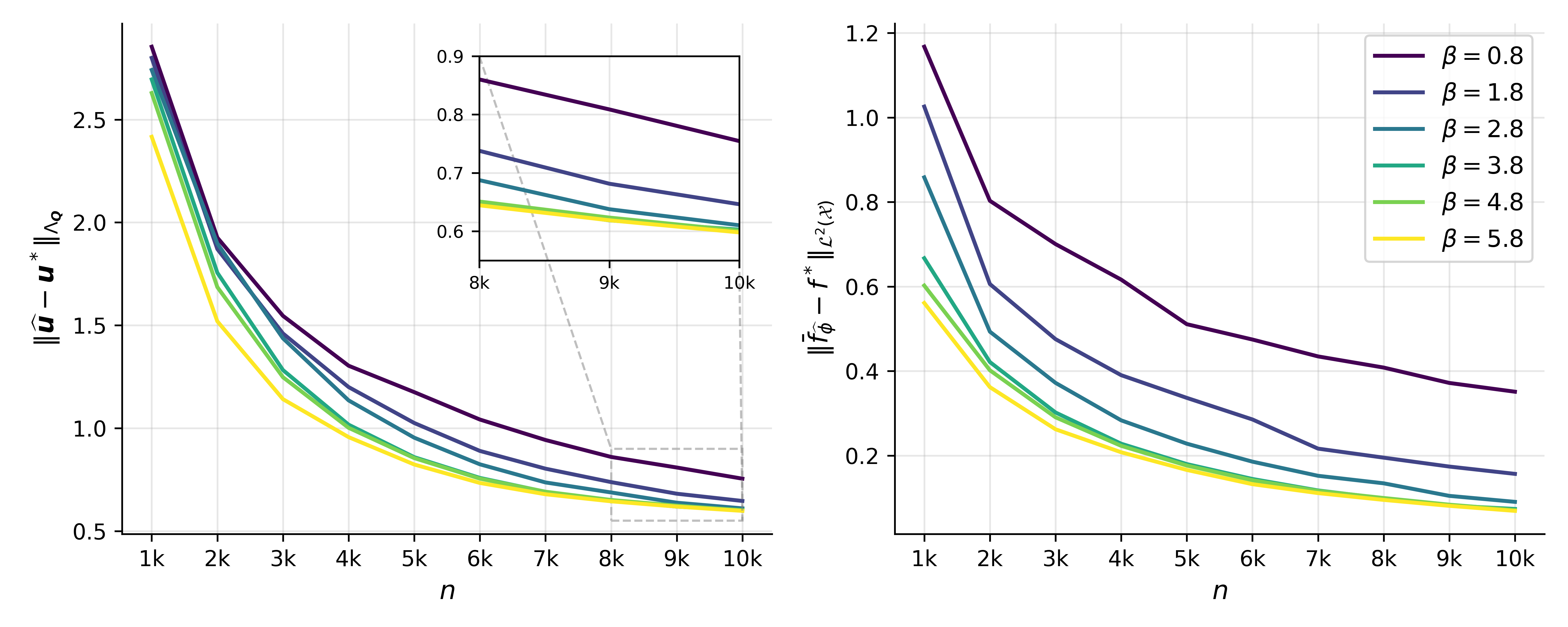}
    \caption{Trace plot of $\| \widehat{\boldsymbol{u}} -\boldsymbol{u}^{*} \|_{\Lambda_{\boldsymbol{Q}}}$ and 
    $\|\bar{f}_{\widehat{\phi}}-f^{*} \|_{\mathcal{L}^{2}(\mathcal{X})}$ as $n$ growing under smoothness parameter $\beta$ of $f^*$. Results are averaged over 300 replications.}
    \label{fig:TDC2}
\end{figure}

To see how the convergence is affected by varying smoothness of the true function $f^*$, we consider the truncated Weierstrass-type function $\mathcal{W}_{\beta}(x) := -\sum_{t=0}^{50} 2^{-t\beta} \left( \cos(2^t \pi x) \right),$ where $\beta \in \mathbb{R}^+$ directly controlling the $\beta$-Hölder smoothness parameter as in Definition \ref{holder_class}.
Then, adding $\mathcal{W}_{\beta}(x_1)+\mathcal{W}_{\beta}(x_2)$ to \eqref{sim1} allows us to control the regularity of the target function. We set the width  in proportion to $(\lfloor\beta\rfloor+1)^2 d^{\lfloor\beta\rfloor+1}$ and depth in proportion to $(\lfloor\beta\rfloor+1)^2\lceil N^{d/(2d+4\beta)} \log _2(8 N^{d/(2d+4\beta)})\rceil$, as those specified in Theorem \ref{thm:consistency}. We examine the model performance under different $\beta$ that are set from $\{0.8,1.8,\ldots,5.8\}$ while fixing $(1-\alpha)=0.15$. The results are presented in Figure \ref{fig:TDC2}.

Overall, the estimation errors are non-increasing when increasing the smoothness of the target function $f^*$.  We also observe a stratification (highlighted by the inset plot) where the marginal improvement in estimation error saturates as $\beta$ becomes sufficiently large. This qualitatively matches the trade-off suggested by Theorem \ref{thm:consistency}: when the function is sufficiently smooth, the nonparametric estimation error becomes negligible compared to the $\widetilde{\mathcal{O}}(\sqrt{n/N})$ term associated with the utility vector estimation.

\subsection{ATP Tennis Data}\label{sec:ATP}

The sports analytics platform Sofascore\footnote{See \url{https://www.sofascore.com/tennis}.} contains professional tennis tournament records of all levels (ATP 250, 500 and 1000, ATP Finals, and Grand Slams) between 2016 and 2025. We collect available information from this platform to construct the dataset, comprising $N=21,385$ professional singles matches among  $n=389 $ players across all ATP tournament tiers (ATP 250, 500, 1000, Finals, and Grand Slams) between 2016 and 2025. Each match is engineered into a 66-dimensional covariate vector, with features divided into three semantic groups outlined in Table~\ref{tab:feature_groups}. We employ a strict temporal split: training on April 2016-June 2023, hyperparameter tuning on June 2023-June 2024, and out-of-sample evaluation on the held-out period June 2024-July 2025.

\begin{table}[!ht]
\centering

\caption{Feature groups for the Sofascore ATP tennis dataset.}
\begin{tabular}{ 
    >{\raggedright\arraybackslash}m{0.22\linewidth} 
    >{\raggedright\arraybackslash}m{0.57\linewidth} 
    m{0.12\linewidth} 
}
\toprule
Feature group & Brief description & \# features \\
\midrule

Match information (MI) & 
Tournament metadata, such as level, surface, and round. & 
\hfill 17 \\
\addlinespace 

Player profile (PP) & 
Per-player identifiers and bio, including age, height, handedness, home advantage, seed rank, \textit{etc}. & 
\hfill 15 \\
\addlinespace

Historical technical statistics (TS) & 
Recent 3-month records including serve/return metrics (aces, double faults, 1st/2nd-serve return points, \textit{etc}), games metrics, tiebreaks, \textit{etc}. & 
\hfill 34 \\
\bottomrule
\end{tabular}
\label{tab:feature_groups}
\end{table}
We compare the DHR model with three baselines. In addition to the BT model introduced above, we include \textbf{PlusDC} 
\citep{dong2025statistical}, which adopts a linear specification $s_{e,j}(\boldsymbol{u}, \boldsymbol{v})=u_j+X_{e,j}^\top 
\boldsymbol{v}$ optimized via the alternating maximization strategy described therein, and an \textbf{Ablated} model $f_{\widetilde{\phi}} \in \max_{f_\phi \in \mathcal{F}_{\mathsf{DNN}}} l(\boldsymbol{0}, f_{\phi})$, which forces the neural network to learn without relying on the intrinsic utility to demonstrate the necessity of heterogeneous effects. To account for randomness in model initialization and training, all models are evaluated over 30 repeated runs with different random seeds. We assess out-of-sample performance by three metrics. The test log-likelihood is computed via \eqref{eq:mle}. Prediction accuracy is the proportion of correctly predicted winners, where the predicted winner on $e_i$ is $$\widehat{\pi}_{e_i}(1)= \underset{j\in e_i}{\operatorname{argmax}} \;\;\widehat{u}_j+\bar{f}_{\widehat{\phi}}(X_{e, j}), \quad \text{and} \quad \widehat{\pi}_{e_i}(2)= e_i \; \backslash \{\widehat{\pi}_{e_i}(1)\},$$ The Brier score is defined as the mean squared difference between the predicted probabilities and the actual outcomes: $$\operatorname{Brier} = \frac{1}{N}\sum_{i \in [N]} \left(\sigma(s_{e_i,j_{i1}}(\widehat{\boldsymbol{u}}, \bar{f}_{\widehat{\phi}})-s_{e_i,j_{i2}}(\widehat{\boldsymbol{u}}, \bar{f}_{\widehat{\phi}})) - \1_{\displaystyle\{ j_{i1}=\pi_{e_i}(1)\}} \right)^2,$$ where $\sigma(\cdot)$ is the sigmoid function. Further details on data processing, feature engineering and other experimental implementations are described in Section E of the Supplement.

\begin{figure}[!ht]
   \begin{subfigure}[b]{0.5\textwidth}
   \centering
   \scriptsize
    \renewcommand{\arraystretch}{2}
    \setlength{\tabcolsep}{2pt}
    \begin{tabular}{lccc}
        \toprule
        & \multicolumn{3}{c}{Evaluation Metrics} \\ 
        \cmidrule(lr){2-4} 
        Model & Accuracy (\%) & Log-likelihood & Brier \\ 
        \midrule
        DHR    & $\mathbf{64.565_{\pm 0.337}}$ & $\mathbf{-0.625_{\pm 0.001}}$ & $\mathbf{0.218_{\pm 0.001}}$ \\
        Ablated & $63.824_{\pm 0.405}$ & $-0.633_{\pm 0.002}$ & $0.222_{\pm 0.001}$ \\
        PlusDC  & $63.351_{\pm 0.000}$ & $-0.647_{\pm 0.000}$ & $0.226_{\pm 0.000}$ \\
        BT      & $60.669_{\pm 0.000}$ & $-0.671_{\pm 0.000}$ & $0.236_{\pm 0.000}$ \\ 
        \bottomrule
    \end{tabular}
    \vspace{0.8cm}
    \caption{}
    \label{fig:atp_level}
    \end{subfigure}
  \hfill 
  \begin{subfigure}[b]{0.45\textwidth}
    \centering
    \includegraphics[width=\linewidth]{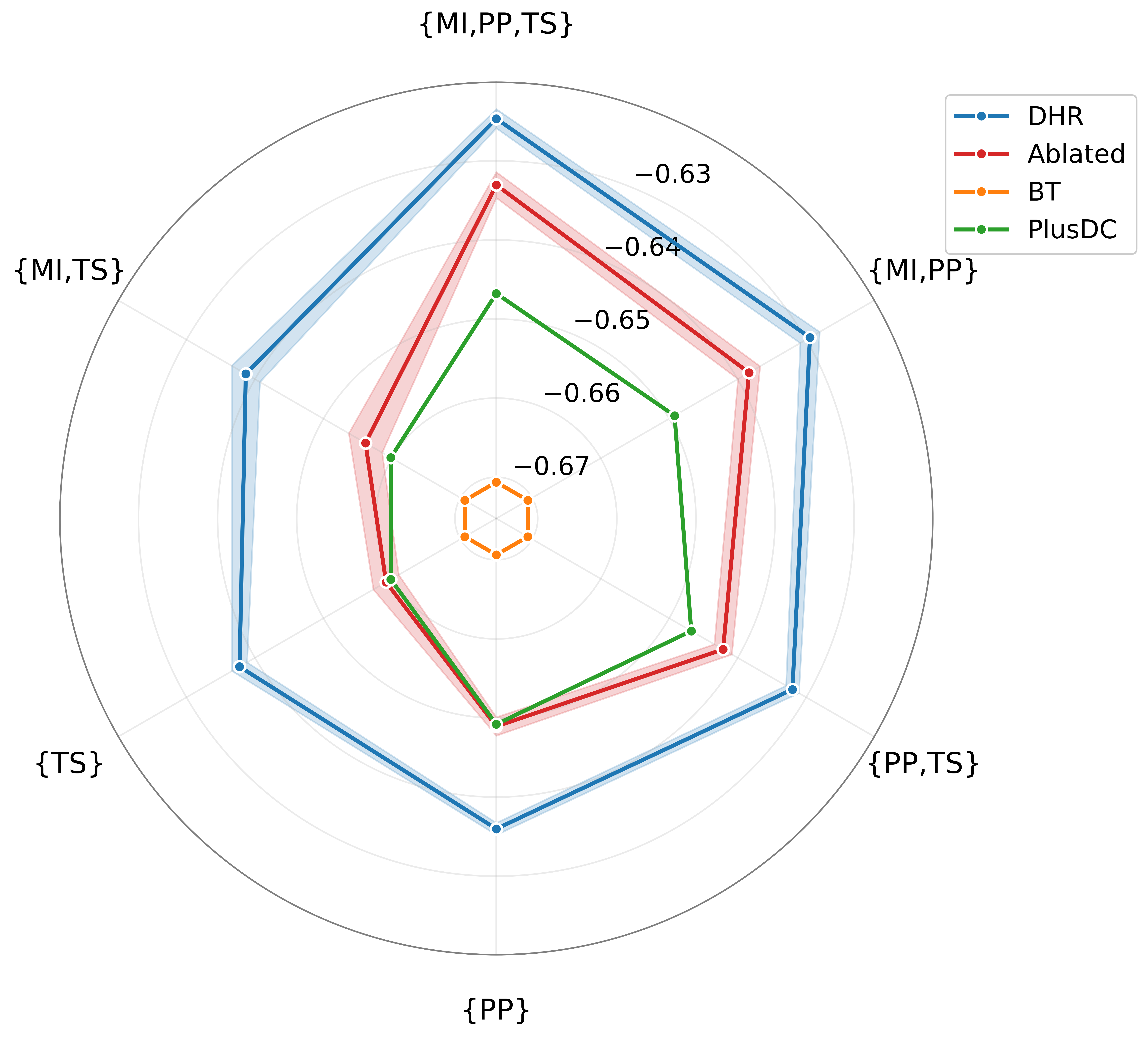}
    \caption{}
    \label{fig:radar_ll}
  \end{subfigure}
  \caption{Test performance of DHR versus baselines. (\subref{fig:atp_level}) Comparison of three metrics on $\{\operatorname{MI}, \operatorname{PP}, \operatorname{TS}\}$ across all models. (\subref{fig:radar_ll}) Radar chart of test log-likelihood across different feature combinations (e.g. from $\{\operatorname{PP}\}$ to $\{\operatorname{MI}, \operatorname{PP}, \operatorname{TS}\}$). The shade represents one standard deviation over 30 replications with different random seeds.}
  \label{fig:oosf}
\end{figure}

\subsubsection{Out-of-Sample Forecasting}\label{oosf}

We first assess the out-of-sample generalization of DHR against other competing methods and examine how it varies under different feature combinations. 

Figure~\ref{fig:oosf}(\subref{fig:atp_level}) reports the out-of-sample performance of DHR and three baseline models on the full feature set. DHR achieves the highest predictive accuracy of $64.565\%$ compared to $63.351\%$ for the PlusDC, $63.824\%$ for the Ablated model, and $60.669\%$ for the BT baseline. Similarly, DHR achieves the lowest Brier score ($0.218$) and the best test log-likelihood ($-0.625$). The Ablated model consistently outperforms both PlusDC and the BT model across all three metrics. This suggests that nonlinear approximation is essential for capturing the complex dynamics of contextual effects in practice. On the other hand, the noticeable gap between DHR and the Ablated model shows that explicitly modeling player-specific heterogeneity is necessary for superior predictive performance.

Figure~\ref{fig:oosf}(\subref{fig:radar_ll}) extends this analysis by examining how log-likelihood varies across different feature combinations. DHR consistently achieves the highest test log-likelihood, maintaining a clear performance margin over all baseline models. Similar patterns are observed under prediction accuracy and Brier score (see Section E of the Supplement).

\begin{figure}[htb!]
    \centering
    \begin{subfigure}{\linewidth}
        \centering
        \includegraphics[width=\linewidth]{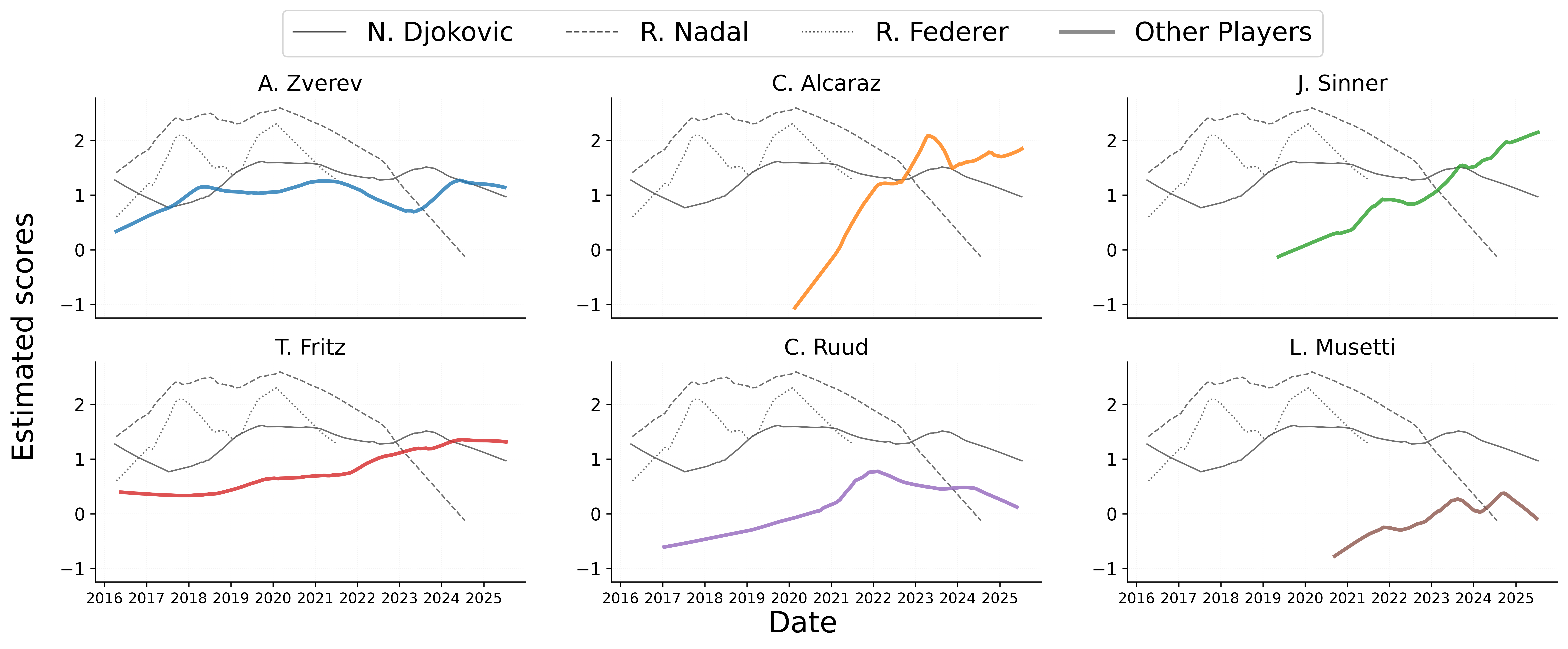}
        \caption{}
        \label{subfig:dynamic}
    \end{subfigure}
    
    
    \begin{subfigure}{\linewidth}
        \centering
        \includegraphics[width=\linewidth]{yearly_bar_deep.png}
        \caption{}
        \label{subfig:cluster_bar}
    \end{subfigure}
    
    \caption{Player scores estimated by DHR. (\subref{subfig:dynamic}) Time-varying trajectories of selected players against the ``Big Three'' benchmarks. (\subref{subfig:cluster_bar}) Average estimated scores across temporal periods. {Error bars represent one standard deviation within each period.}}
    \label{fig:timely_journey}
\end{figure}

\subsubsection{Temporal Evolution of Player Estimated Scores}

Another aspect of sports analytics lies in understanding the individualized dynamics of players' performance in their careers. We examine this by analyzing the time-varying trajectories derived by $(\widehat{\boldsymbol{u}},\bar{f}_{\widehat{\phi}})$. For a player $j$, the trajectory is constructed by temporally ordering the estimated scores $\{ \widehat{u}_j + \bar{f}_{\widehat{\phi}}(X_{e_i,j}) \}_{e_i \in \mathcal{E}_n}$.  We use the familiar ``Big Three'' (N. Djokovic, R. Nadal, and R. Federer) as reference baselines to assess how other players' estimated scores evolve relative to them over time.  Figure~\ref{fig:timely_journey} presents the estimated time-varying trajectories of selected players from a single run using a fixed random seed. In the upper panel~(\subref{subfig:dynamic}), these trajectories are smoothed via LOWESS\footnote{LOWESS smoothing is implemented via \texttt{statsmodels} package in Python.} 
for visualization, while bottom panel~(\subref{subfig:cluster_bar}) reports period-wise average estimated scores.

During 2016–2021, the ``Big Three'' generally maintained a high level of competitiveness, which is reflected by their estimated trajectories occupying the upper range in Figure~\ref{fig:timely_journey}(\subref{subfig:dynamic}) and the corresponding averages in Figure~\ref{fig:timely_journey}(\subref{subfig:cluster_bar}). Among them, N. Djokovic exhibited a relatively stable estimated trajectory over this period. The estimated trajectory of R. Nadal showed declines in the later years. Among the younger generation, A. Zverev, who won the ATP Finals in both 2018 and 2021 and reached the 2020 US Open final, maintained a relatively stable estimated trajectory near or below the ``Big Three'' throughout this period. By contrast, other younger players such as T. Fritz, C. Alcaraz, J. Sinner, and C. Ruud, all of whom were still in the early stages of their careers, still showed a gap from the level of the ``Big Three'', consistent with the lower average scores reported in Figure~\ref{fig:timely_journey}(\subref{subfig:cluster_bar}) for this period.

From 2021 onward, a noticeable shift in the estimated score hierarchy began to emerge. Notably, C. Alcaraz and J. Sinner won multiple Grand Slam titles. C. Alcaraz rose to World No.~1 in 2022 and Sinner reached World No.~1 in 2024. This rapid rise is captured by their most pronounced upward trajectories in Figure~\ref{fig:timely_journey}(\subref{subfig:dynamic}), where they gradually close the gap with the ``Big Three'' and eventually surpass them around 2022-2023. This shift is further observed in Figure~\ref{fig:timely_journey}(\subref{subfig:cluster_bar}), where both players achieve the highest average estimated scores in the last two periods. Similarly, C. Ruud showed marked improvement and reached an epic of his career in 2022, finishing runner-up at both the 2022 French Open and the 2022 US Open. This milestone is reflected in his estimated trajectory around 2022 in Figure~\ref{fig:timely_journey}. 
A. Zverev and T. Fritz remained strong and stable competitors over this period. A. Zverev reached the 2024 French Open final but was defeated by C. Alcaraz, while T. Fritz advanced to the 2024 US Open final, ultimately losing to J. Sinner.  Their estimated trajectories reflect this gap, remaining at a high level but falling short of the peak levels reached by C. Alcaraz and J. Sinner. L. Musetti exhibited gradual improvement as well, but his estimated scores mostly remained below those of other selected players.

Overall, the log-scores of the players over time based on their respective covariate information, as uncovered by DHR are consistent with their empirical performance observed in practice. These results reveal a dynamic shift from the dominance of the ``Big Three'' in the earlier period to a more competitive landscape after 2021.

\section{Discussion}
In this work, we have introduced a semiparametric framework for statistical ranking that integrates the interpretability of PL models with the nonlinear adaptability of deep neural networks. By modeling the log-score as an additive combination of an object-specific intrinsic parameter $u_j$ and covariate effects $f(X_{e,j})$, we address the limitations of traditional parametric models that struggle to disentangle utility from dynamic contextual advantages.

While we established consistency in the graph-weighted norm, deriving a uniform $\ell_\infty$ bound is complicated by the $\mathcal{L}^\infty(\mathcal{X})$ estimation error of the nonparametric component, which may require new techniques from DNN estimation theory.  Alternatively, another promising direction is to investigate whether refinement procedures, such as those proposed by \citet{chen2024anote}, can be applied to establish $\ell_{\infty}$ error control and extend the theory to sparser graph regimes. Beyond estimation errors, ranking inference for quantities of interest is another important topic \citep{yan2025likelihood,fan2025spectral,han2025unifiedanalysis}. However, establishing the asymptotic normality of the estimators in this setting is presumably challenging.

Since our framework provides broad methodological flexibility, a natural extension would be to accommodate multi-level outcomes, such as ties or Likert-scale responses. Furthermore, while this work focused on standard DNNs, our nonparametric estimation procedure is agnostic to the chosen function class. Consequently, future implementations could incorporate neural additive models \citep{agarwal2021neural} to enhance interpretability, or utilize transformer architectures \citep{vaswani2017attention} to capture more complex covariate representations.

\begin{description}
	\item[Supplementary material:] We include detailed proofs of Theorems \ref{lemma_ident}--\ref{thm:consistency} and Proposition \ref{lemma_uni_bounded}. In addition, we provide additional numerical results and detailed procedures for ATP tennis data analysis.
	\vspace{-0.3cm}
	\item[Data and Code:] We include the data and Python code to reproduce the results in numerical studies, see \url{https://github.com/fsmiu/Deep-Ranking-with-Heterogeneous-Effects}.
\end{description}

\spacingset{1.6} 

\bibliography{reference.bib}
\end{document}